\documentclass[11pt]{article}
 \usepackage{geometry}
 \usepackage[affil-it]{authblk}
 
 \usepackage[square,authoryear]{natbib}

 \usepackage{amsmath}
 \usepackage{amssymb}

\usepackage{graphicx}

\usepackage{caption}
\usepackage{subcaption}

\usepackage{epsfig}
\usepackage{hyperref}
\usepackage{epstopdf}
\usepackage{color}
\def\beq{\begin{equation}}
\def\be{\begin{equation}}
\def\ee{\end{equation}}
\def\bes{\begin{eqnarray}}
\def\ees{\end{eqnarray}}

\title{Mapping systemic risk: critical degree and failures distribution in financial networks}

\author{Matteo Smerlak\footnote{Corresponding author. Email: msmerlak@perimeterinstitute.ca. Telephone: (+1) 226-339-2614.}}\affil{Perimeter Institute for Theoretical Physics, 31 Caroline Street North, N2L 2Y5 Waterloo ON, Canada}

\author{Brady Stoll}\affil{Department of Mechanical Engineering, University of Texas at Austin, 204 E. Dean Keaton, C2200, Austin, TX 78712, USA}

\author{Agam Gupta}\affil{Indian Institute of Management, Diamond Harbour Road
Joka, Kolkata (Calcutta), 700104
West Bengal, India}

\author{James S. Magdanz}\affil{Resilience and Adaptation Program,
University of Alaska Fairbanks,
PO Box 757000,
Fairbanks, AK 99775-7000, USA}

\begin{document}

\maketitle







\section*{Abstract}

The 2008 financial crisis illustrated the need for a thorough, functional understanding of systemic risk in strongly interconnected financial structures. Dynamic processes on complex networks being intrinsically difficult, most recent studies of this problem have relied on numerical simulations. Here we report analytical results in a network model of interbank lending based on directly relevant financial parameters, such as interest rates and leverage ratios. Using a mean-field approach, we obtain a closed-form formula for the ``critical degree", viz. the number of creditors per bank below which an individual shock can propagate throughout the network. We relate the failures distribution (probability that a single shock induces $F$ failures) to the degree distribution (probability that a bank has $k$ creditors), showing in particular that the former is fat-tailed whenever the latter is. Our criterion for the onset of contagion turns out to be isomorphic to the condition for cooperation to evolve on graphs and social networks, as recently formulated in evolutionary game theory. This remarkable connection supports recent calls for a methodological rapprochement between finance and ecology.

%
%


\section{Introduction}

In the financial sector, shock propagation mechanisms are at the core of systemic risk \citep{DeBandt:2000vm,Allen:2009gl}, and banks play the most important role \citep{Billio:2012vj}. An important and potentially vulnerable arena for financial contagion is the interbank loan market, which allows banks to rapidly exchange large amounts of capital for short durations to accommodate temporary liquidity fluctuations \citep{Furfine:1999ho,Ashcraft:2007vf,Taylor:2008ux}. Consequently, interbank loan networks have been of particular interest in exploring systemic risk  \citep{Gai:2010tr,May:2010ev,Acemoglu:2013ui}. 

In recent years, random network theory \citep{Newman:2010wp,Barrat:2008wp} has provided a useful framework to study cascade effects in interconnected structures \citep{Watts:2002kb}. Applied to the financial sector \citep{May:2008uk}, network approaches have clarified the role of connectivity  \citep{Nier:2007cc,Battiston:2012vu}, bank size \citep{Arinaminpathy:2012uk}, shock size \citep{Acemoglu:2013ui} and overlapping portfolios \citep{Caccioli:2012vq} in systemic risks. Increased understanding of contagion in finance \citep{Aghion:2000bm,Furfine:2003wl} has led to an increased interest by regulators and central bankers \citep{Weidman:2007us,Bisias:2012jl} in using network measures to evaluate systemic risk.
 
 An essential insight of Allen and Gale \citep{Allen:2000er}, confirmed in \citep{Gai:2010tr} and deepened in \citep{Battiston:2012vu}, is that increasing network connectivity---measured by its mean degree $z$---can have opposite effects depending on the baseline value of $z$. On the one hand, when the network is sparsely connected, increasing $z$ will open new channels for contagion and weaken the network. On the other hand, when $z$ is sufficiently large, increasing $z$ further will dilute the effect of a localized shock and strengthen the network. From this perspective, the key question is not \emph{if}, but \emph{when}, enhanced connectivity helps secure network robustness.  

Our first goal in this paper is to sharpen these results by introducing a model of interbank lending that allows the ``critical degree" separating these two regimes to be computed as an explicit function of a small number of relevant financial parameters: (interbank and external) interest rates, liquidity requirement, leverage ratio. As we shall see, this critical degree is pivotal in deriving an analytical estimate of the number of failures induced by a single shock given these parameters. Our results complement those of \citep{Gai:2010tr}, who used the mathematics of percolation theory \citep{Newman:2001un} to determine the contagion threshold in financial networks, as well as those of \citep{May:2010ev}, who brought to bear the ``mean-field approximation'' familiar to statistical physicists.

Our second goal is to analyze the role of degree heterogeneity in financial networks with regard to systemic risk. It has long been known \citep{Albert:2000bc,Newman:2002jj} that network topology is a key determinant of network robustness. Empirical studies of flows over the Fedwire Funds Services \citep{Soramaki:2007jc,Bech:2010vk} have found the network to be inhomogeneous, with a strongly connected, strongly reciprocal core and a much more sparse periphery.\footnote{Similar analyses have been conducted of interbank loan networks in Belgium \citep{Degryse:2007wo}, Austria \citep{Boss:2004hv}, the Netherlands \citep{Propper:2008tu}, Italy \citep{Mistrulli:2007tf}, and East Asia \citep{Inoguchi:2013ts}, with similar results.} Nonetheless, most theoretical studies of the systemic risk to date \citep{Allen:2000er,May:2010ev,Gai:2010tr} have used homogeneous (Erd\"os-R\'enyi) networks. We show that, when banks' degrees have a fat-tailed distribution, the number of failures induced by a single shock follows a similar distribution---a precise statement of the ``robust-yet-fragile'' property of financial networks emphasized by several authors \citep{Haldane:2009uv,Gai:2010tr,Acemoglu:2013ui}. 

The paper is organized as follows. We begin by describing our model of interbank lending networks, first in some generality and then under simplifying assumptions. Next we show how the number of failures induced by an individual shock can be estimated analytically by means of a mean-field-type approximation, in which Cayley trees (regular networks without loops) play an instrumental role. We then compare our results with numerical simulations of both homogenous and scalefree random networks. We close with a few remarks concerning the policy implications of our work, and point out an intriguing biological analogy.

\section{Model}

\subsection{Interbank network}

We present a model\footnote{A preliminary analysis of this model was reported in \citep{Gupta:2013vx}.} of the structure of interbank lending as a random weighted directed network, in which a node $i\in\{1,\cdots,N\}$ represents a bank and a link $i\rightarrow j$ with weight $l_{ij}$ a loan of amount $l_{ij}$ made by $i$ to $j$. The sum of all weights flowing out of a bank $i$, $l_{i}=\sum_{j\leftarrow i}l_{ij}$, is therefore the total interbank exposure of bank $i$; the sum of weights flowing into $i$, $b_{i}=\sum_{j\rightarrow i}l_{ji}$, is in turn the total liability of bank $i$ on the interbank market.

\subsection{Balance sheets}

In addition to its interbank liabilities $b_{i}$, we assume that each bank $i$ has external, more senior liabilities $s_{i}$ (e.g. deposits). On the asset side, we further introduce liquid assets $\lambda_{i}$ (e.g. bonds) as well as illiquid assets $\iota_{i}$ (e.g. buildings). The total assets $A_{i}$ and total liabilities $L_{i}$ of bank $i$ can therefore be written as $A_{i}=l_{i}+\lambda_{i}+\iota_{i}$ and $L_{i}=b_{i}+s_{i}$; the difference $K_{i}=A_{i}-L_{i}$ is the net worth of bank $i$, see Table \ref{sheets}. 

\bigskip

\begin{table}[h]
 \begin{center}
\begin{tabular}{@{\vrule height 10.5pt depth4pt  width0pt}|c|c|}
  \hline
  assets $A_{i}$ & liabilities $L_{i}$ \\
  \hline
 liquid assets $\lambda_{i}$ & senior liabilities $s_{i}$\\
 illiquid assets $\iota_{i}$ & interbank borrowings $b_{i}$\\ \cline{2-2}
  interbank loans $l_{i}$ & net worth $K_{i}$\\
  \hline
\end{tabular}
\end{center}
\caption{Balance sheet of bank $i$.}
\label{sheets}
\end{table}

Basel III \citep{Committee:2010vl} introduced leverage and liquidity requirements for banks. We define for each bank $i$ the \emph{leverage ratio} $\Lambda_{i}=K_{i}/A_{i}$ (ratio of networth to total assets) and the \emph{liquidity ratio} $f_{i}=\lambda_{i}/A_{i}$ (ratio of liquid assets to total assets). By definition, lowering the ratios $\Lambda_{i}$ and $f_{i}$ increases the exposure of bank $i$ on the interbank market; we shall see that they have a strong impact on the systemic risk.

\subsection{Repayment equation}

We now introduce a two-period investment dynamics, through which a bank can either increase or decrease its net worth $K_{i}$. In the first step, a bank uses its total liabilities $L_{i}$ to invest in some external opportunity, at some interest rate $R_{i}$. (Successful investments correspond to $R_{i}>1$, hazardous ones correspond to $R_{i}<1$; in the worst case scenario, the investment is lost in full, viz. $R_{i}=0$.) We denote $\rho_{i}=(R_{i}-1)L_{i}$ the profit made in this transaction.\footnote{If $s_{i}<0$, we take $\rho_{i}=(R_{i}-1)b_{i}$; equivalently, the profit is defined by $\rho_{i}=(R_{i}-1)\max\{\textrm{liab}_{i},b_{i}\}$.} In the second step, a bank uses this profit and its liquid assets to repay its interbank liabilities $l_{i}$ with an interest $r>1$ while ensuring the seniority of $s_{i}$. 

Denoting $x_{ij}$ the amount repaid by bank $i$ to bank $j$ in the second step, we assume the following repayment rules.\footnote{Strictly related rules, inspired from Einsenberg and Noe's seminal paper \citep{Eisenberg:2001uoa}, were used recently in \citep{Acemoglu:2013ui}.}

\begin{itemize}
\item
\emph{Full repayment}: if $\rho_{i}+\lambda_{i}-s_{i}+\sum_{j\neq i}x_{ji}\geq rb_{i}$, bank $i$ repays its junior debt $rb_{i}$ in full, hence for each $j\neq i$
$$x_{ij}=rl_{ji},$$ 
\item
\emph{Partial default}: if $0<\rho_{i}+\lambda_{i}-s_{i}+\sum_{j\neq i}x_{ji}<rb_{i}$, bank $i$ repays a fraction of its junior liabilities on a \emph{pro rata} basis, hence for each $j\neq i$ $$x_{ij}=\frac{l_{ji}}{b_{i}}\left(\rho_{i}+\lambda_{i}-s_{i}+\sum_{j\neq i}x_{ji}\right)$$
\item
\emph{Complete default}: if $\rho_{i}+\lambda_{i}-s_{i}+\sum_{j\neq i}x_{ji}\leq 0$, bank $i$ repays nothing, hence $x_{ij}=0$ for each $j\neq i$. 
\end{itemize} 
When a bank $i$ can just repay its interbank borrowings, i.e. when
\be
\rho_{i}+\lambda_{i}-s_{i}+\sum_{j}(l_{ij}/b_{i})x_{j}=rb_{i},
\ee
we say that $i$ is \emph{critical}. After all repayments are made, bank $i$ has an updated net worth
\be\label{networth}
K'_{i}=\rho_{i}+\lambda_{i}+\iota_{i}-s_{i}+\sum_{j\neq i}(x_{ji}-x_{ij}).
\ee
We call ``safe'' the banks $i$ such that $K'_{i}>0$, and ``failed'' the ones such that $K'_{i}\leq0$. 

\subsection{Simplifying assumptions}

From a mathematical perspective, finding the interbank repayments involves solving the system of $N$ coupled, non-linear equations
\be\label{repayment}
x_{i}=\left[\min\Big\{\rho_{i}+\lambda_{i}-s_{i}+\sum_{j\leftarrow i}(l_{ij}/b_{i})x_{j},rb_{i} \Big\}\right]^{+},
\ee
where $[\,\cdot\,]^{+}=\max\{\,\cdot\,,0\}$ and the sum ranges over $i$'s debtors; the repayment $x_{ij}$ of bank $i$ to bank $j$ is then given by $x_{ij}=l_{ij}x_{i}/b_{i}$. 

While the set of equations \eqref{repayment} can be studied numerically for various network topologies and different values of the financial parameters $R_{i}$, $f_{i}$, $\Lambda_{i}$ and $r$, our goal in this paper is to obtain explicit, analytical results about the robustness of financial networks with respect to shocks. To make progress, we make the following---dramatic but empowering---assumptions: $(i)$ all loans are reciprocated, so that the network is actually \emph{undirected},\footnote{As observed in Ref. \citep{Soramaki:2007jc}, the reciprocity of the ``core'' financial network is very high, making this assumption less unrealistic than may seem at first sight.} $(ii)$ all interbank loans $l_{ij}$ have unit value, so that $l_{i}=b_{i}=k_{i}$, where $k_{i}$ is the degree of node $i$, $(iii)$ all banks have equal leverage and liquidity ratios $(\Lambda,f)$, so that the latter can be thought of as model parameters rather than individual variables $(iv)$ illiquid assets are negligible ($\iota_{i}=0$), and $(v)$ external interest rates $R_{i}$ take the same value $R>1$ for all banks across the network except one (bank $i=i_{0}$, call it the ``shocked bank''), for which $R_{i_{0}}=0$. 

Within this simple setting, our objective is then to estimate the ``number of induced failures'' $F$ (defined as the number of banks $i\neq i_{0}$ such that $K_{i}'\leq0$) as a function of the financial parameters $(R,r,\Lambda,f)$ and of the network topology.


\section{Results}

\subsection{Cayley trees}
We begin our investigation of the model by considering the simplest network topology, namely a network with uniform degree $k$ and no loops (a ``Cayley tree''). On such simple networks, the repayment problem \eqref{repayment} can be solved exactly (see Materials and Methods and SI for details), as follows. 

When $k$ is sufficiently large, each neighbor of the shocked bank $i_{0}$ inherits only a small fraction of $i_{0}$'s losses---and none fails. For networks with incrementally decreasing degree $k$, however, the effect of these losses on the net worth of each creditor of $i_{0}$ gradually increases, until at some point shocked bank $i_{0}$'s weakest neighbor also fails. If degree $k$ further decreases, the second (then third, etc.) neighbors of $i_{0}$ also approach criticality, and start failing as well.

This sequence of transitions, involving higher and higher neighbors of the shocked bank, defines an ordered sequence of ``critical degrees'' $k_{(1)}^{*}>k_{(2)}^{*}>...$ such that
\be
F=\sum_{p=1}^{q}N_{(p)}(k)\quad \textrm{for} \quad k_{(q+1)}^{*}<k\leq k_{(q)}^{*}
\ee
where $N_{(p)}(k)=k(k-1)^{p-1}$ is the number of nodes at distance $p$ from $i_{0}$. The values of these critical degrees provide a measure of the robustness of the network with respect to a shock: the higher the critical degrees, the more fragile the financial structure.

\begin{figure}[t!]
 \begin{center}
   \includegraphics[scale=.6]{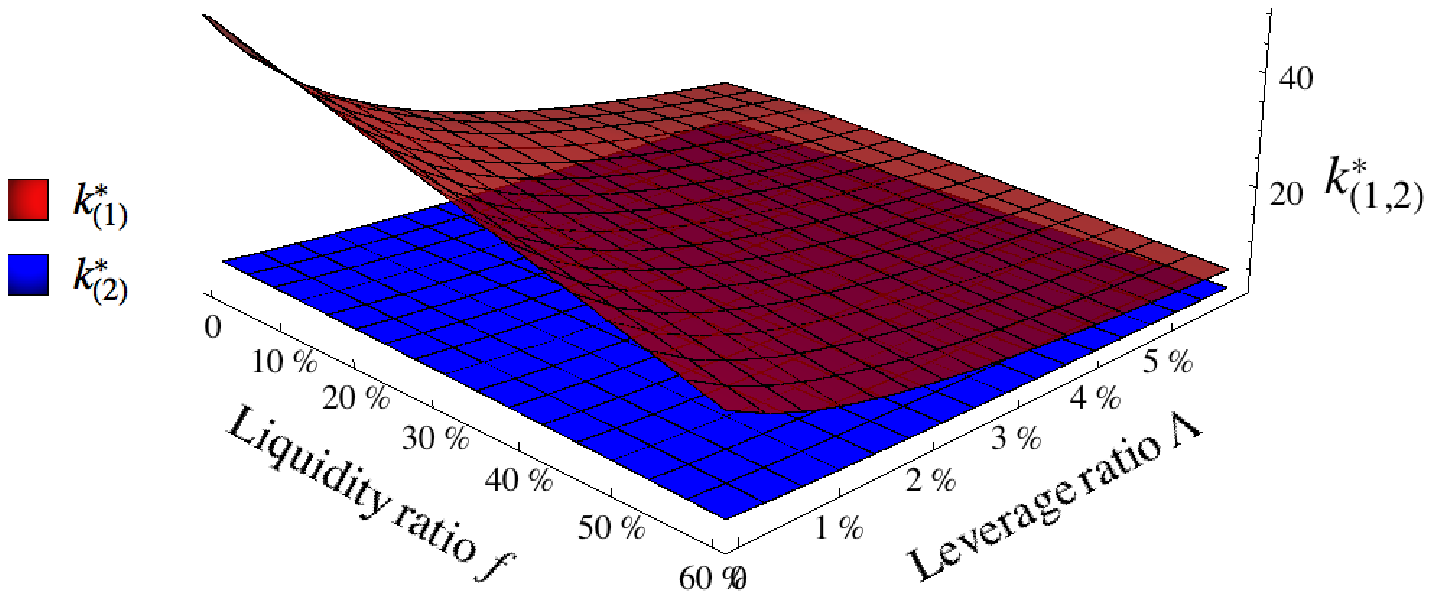}\hfill
   \includegraphics[scale=.6]{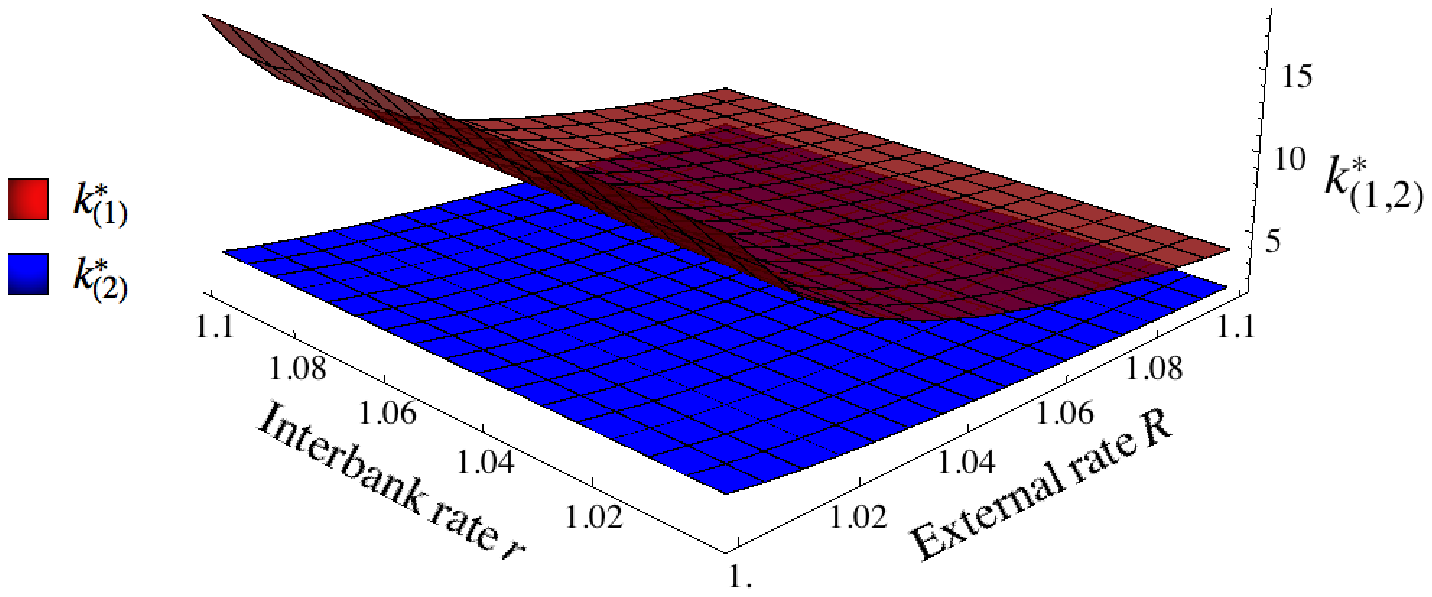}
 \end{center}
 \caption{The first two critical degrees $k_{(1)}^{*}$ and $k_{(2)}^{*}$ as functions of the liquidity ratio $f$ and of the leverage ratio $\Lambda$ for $R=1.05$, $r=1.01$ (left) and as functions of the external rate $R$ and the interbank rate $r$ for $f=50\%$ and $\Lambda=3\%$ (right).}
 \label{figcritical}
 \end{figure}

The expression for each $k_{(q)}^{*}$ as a function of the financial parameters $(R,r,f,\Lambda)$ can be obtained by solving the repayment equations \eqref{repayment} under these conditions that $(i)$ all banks at distance $d\geq q$ from the shocked bank are safe, but $(ii)$ all $q$-th neighbors of $i_{0}$ are critical. This gives in particular
\begin{eqnarray}\label{criticaldegree}
k_{(1)}^{*}&=&\frac{r(1-f)-[r(1-f)+2\Lambda-1]^+}{(R-1)(1-\Lambda)+\Lambda}.
\end{eqnarray}
The critical degrees $k_{(1)}^{*}$ and $k_{(2)}^{*}$ (given explicitly in SI) are plotted as functions of the financial ratios $(f,\Lambda)$ for $R=1.02$ and $r=1.01$ and as functions of the interest rates $(R,r)$ for $f=50\%$ and $\Lambda=3\%$ in Fig. \ref{figcritical}. As is apparent from these plots, $k_{(1,2)}^{*}$ are stricly decreasing functions of $f$ and $\Lambda$: lower liquidity leverage ratios both enhance the systemic risk---an intuitive conclusion, which is here proved rigorously. Furthermore, we see that, unlike the first critical degree $k_{(1)}^{*}$, the second critical degree never becomes appreciably large, $k_{(2)}^{*}\lesssim 5$, so that failures in effect hardly extend beyond the first neighbors of the shocked bank. Finally, we note that, in the limit where $\Lambda,f\rightarrow0$ (a regime in which the economy is dominated by interbank transactions), the first critical degree $k_{(1)}^{*}$ reaches the value $1/(R-1)$; we will come back to this observation in the concluding section.


\subsection{General networks}

Real-world financial networks being anything but regular, the usefulness of the exact solution above would seem to be extremely limited. It turns out to be the opposite. In the regime where failures are unlikely to extend beyond the first neighbors of the shocked bank---which the case for most realistic values of the financial parameters, as illustrated in Fig. \ref{figcritical}---knowing the first critical degree (hereafter denoted simply $k^{*}$) yields a reliable estimate of the failures distribution on random networks, including scalefree ones.

We will assume that, on a general random network, \emph{a first neighbor $i$ of the shocked bank will indeed fail if and only if its own degree $k_{i}$ is smaller than the critical degree $k^{*}$ given by \eqref{criticaldegree}}. This is akin to the ``mean-field'' approximation familiar from statistical mechanics: it replaces the actual, inhomogeneous, environment of $i$ in the network by a homogeneous environment in which all banks have the same degrees as $i$, here the Cayley tree with degree $k_{i}$. While this approximation clearly cannot capture the dynamics of a single network, it does provides a tractable starting point to study the statistics of failures in a given ensemble of random networks. Within this approximation, we obtain the following results (Materials and Methods and SI).

First, the expected number of failures can be estimated as
\be\label{failures}
\langle F\rangle=\sum_{k\geq1}kp(k)q(k)+\cdots,
\ee
where $q(k)=\sum_{l=1}^{k^{*}}p(l\vert k)$ is the probability that a neighbor of the shocked bank $i_{0}$ has a subcritical degree and the dots indicate that the contribution of higher neighbors of $i_{0}$ has been neglected. Here $p(k)$ is the degree distribution (probability that a node has degree $k$) and $p(l\vert k)$ is the conditional degree distribution (probability that a node attached to a node with degree $k$ has degree $l$). Note that formula \eqref{failures} implies that disassortative financial networks (for which the probability $q(k)$ that a neighbor of the shocked bank has subcritical degree increases with the number of neighbors $k$) tend to be more vulnerable to contagion than assortative or uncorrelated ones \citep{Newman:2002jj}.\footnote{In the latter case, one checks that \eqref{failures} reduces to $\langle F\rangle=\sum_{l=1}^{k^{*}}lp(l)=qz$, where $q(k)=q$ is independent of $k$.} 

Second, we show\footnote{Provided the network is asymptotically uncorrelated, \emph{viz}. $p(l\vert k)$ is independent of $k$ when $k\gg l$; see SI.} that, whether the network is Poisson-distributed ($p(k)\sim z^{k}/k!$) or power-law distributed ($p(k)\sim k^{-\gamma}$), the failures distribution has the same asymptotic behavior as the degree distribution itself. In the scalefree case, this means in particular that $P(F)$ has a power-law falloff with exponent $\gamma$, hence is \emph{fat tailed}. This result can be interpreted as expressing the ``robust-yet-fragile'' property of scalefree networks noted earlier: even when the expected number of failures $\langle F\rangle$ is low, the risk remains that a single shock can take down a significant fraction of the network.

\begin{figure}
\centering
  \raisebox{6mm}{\includegraphics[scale=.55]{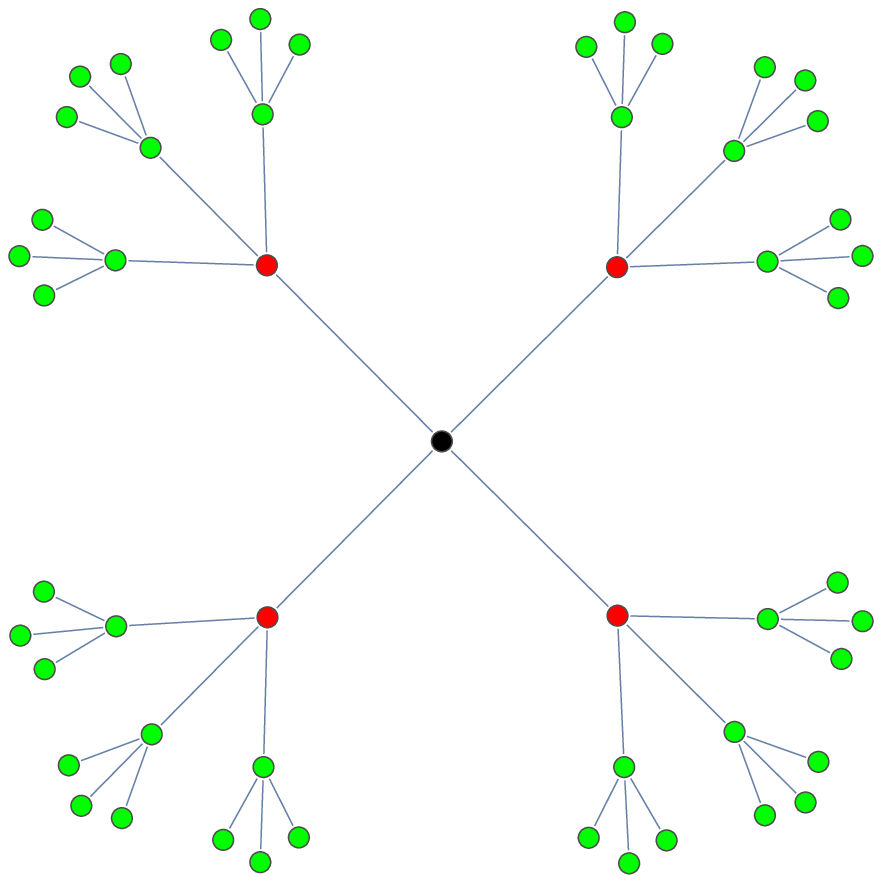}}\hspace{3.5mm}
   \raisebox{10mm}{\includegraphics[scale=.65]{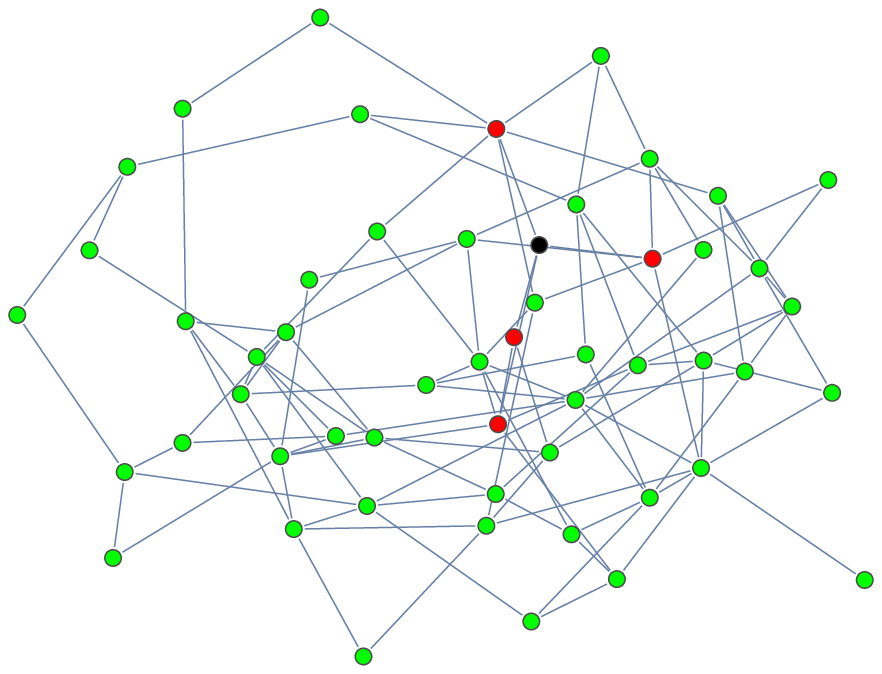}}
   \raisebox{6mm}{\includegraphics[scale=.65]{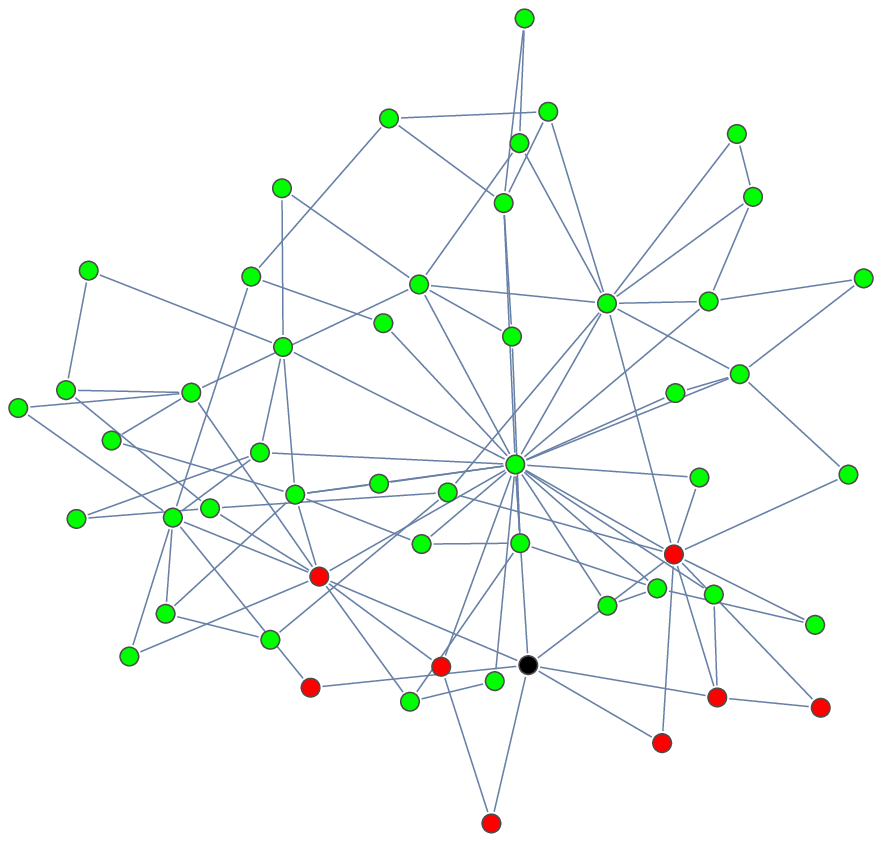}}
 \caption{Failures in sample networks with $N=53$ and $z=4$: from left to right, Cayley tree, ER network, BA network. The black node indicates the shocked bank, the red nodes the failed banks, the green nodes the safe banks. Here $R=1.02$, $r=1.01$, $f=50\%$ and $\Lambda=3\%$.}
 \label{attack}
        \end{figure}

\subsection{Numerical tests}\label{num}

\begin{figure}[t!]
\centering
\begin{subfigure}[b]{\textwidth}
\includegraphics[scale=.9]{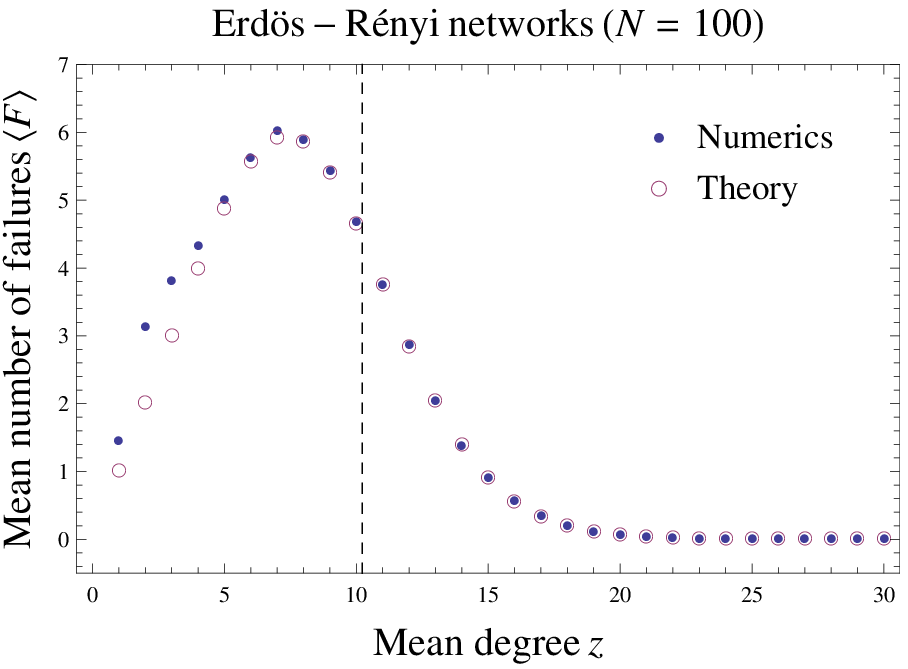} \hfill
    \includegraphics[scale=.9]{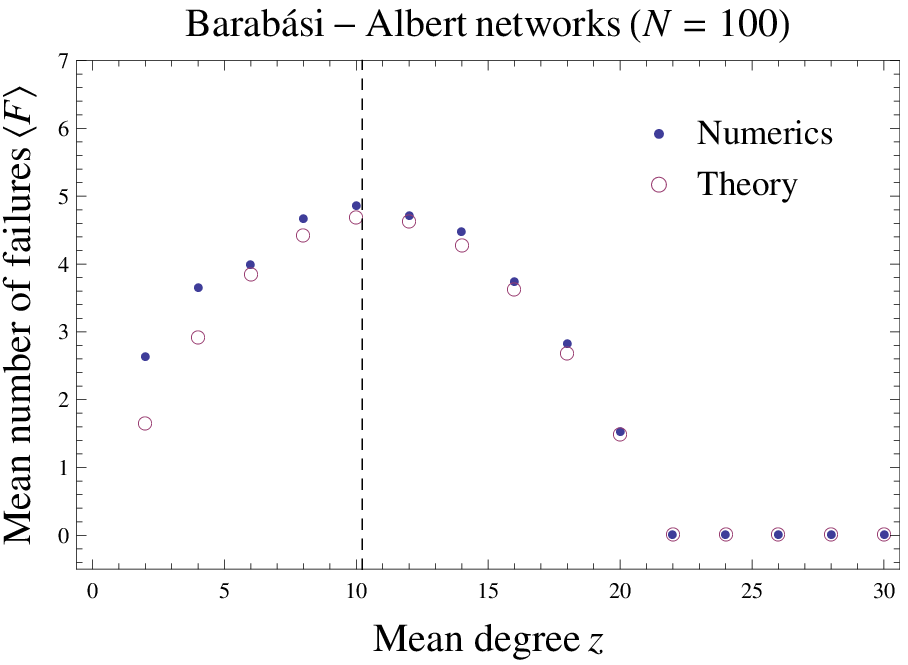} 
 \caption{Mean number of failures as a function of mean degree $z$, as estimated analytically (circles) and as obtained numerically (dots). The dashed line indicates the value of the critical degree $k^{*}$. The discrepancy between the empirical and theoretical values at low $z$ is due to the contribution of second neighbors, neglected in our approximation.}
 \label{res1}
        \end{subfigure}
        \smallskip
       
        \begin{subfigure}[b]{\textwidth}
  \includegraphics[scale=.9]{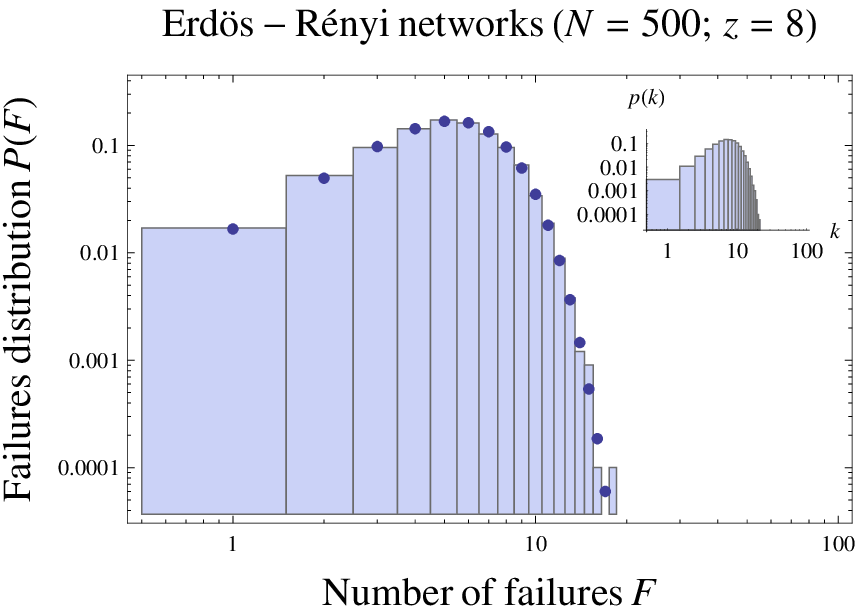}\hfill
    \includegraphics[scale=.9]{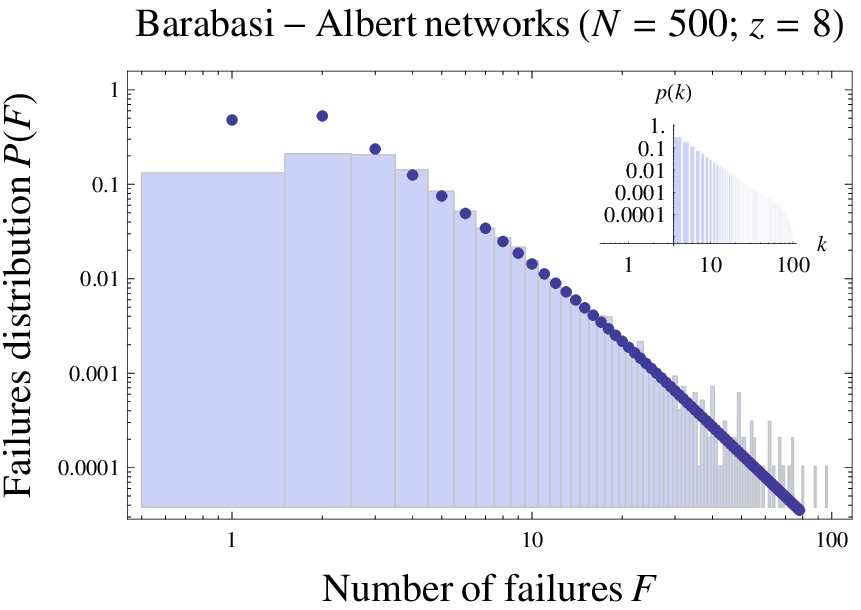}
 \caption{Log-scaled histograms of the empirical (bars) and predicted (dots) failures distribution, for an ensemble of $10^{4}$ ER and BA networks with $N=500$ and $z=8$; the insets show the corresponding degree distributions.}
 \label{res2}
        \end{subfigure}
               \caption{Statistics of failures in ER (left) and BA (right) networks, for $R=1.02$, $r=1.01$, $f=50\%$ and $\Lambda=3\%$. Observe the ``robust-yet-fragile'' nature of scalefree networks: while the maximum expected number of failures is lower than for ER networks, the probability of catastrophic failures is much higher.}
\end{figure}

To test the validity of these findings, we analyzed two types of random networks for which the conditional probability distribution $p(l\vert k)$ is known explicitly as a function of the mean degree $z$ (at least in the large $N$ limit): the classical Erd\"os-R\'enyi (ER) model \citep{Erdos:1959wn}, with Poisson degree distribution, and the Barab\'asi-Albert (BA) model \citep{Barabasi:1999uu}, with scalefree degree distribution $P(k)\sim k^{-3}$. 

For both network types, we generated $10^{4}$ random networks for each value of the mean degree $z$. We solved Eq. \eqref{repayment} numerically for each of these networks and, using Eq. \eqref{networth}, we computed the number of induced failures $F$. From this we determined the mean number of failures and the empirical failures distribution at each $z$, and compared them with our mean-field estimates. Finally, we checked that using directed networks (both random and scalefree) does not yield significantly different results, thereby confirming the validity of assumption $(i)$ as a useful first approximation.

Fig. \ref{res1} shows the mean number of failures $\langle F\rangle$ as a function of the mean degree $z$ for homogeneous ER and scalefree BA networks, in the  parameter regime $R=1.02$, $r=1.01$, $f=50\%$ and $\Lambda=3\%$ (for which $k^{*}\simeq 10.2$); see also Fig. S1. Irrespective of the network topology, we find that the empirical value of $\langle F\rangle$ matches very closely with our estimate \eqref{failures} provided that the mean degree is not too small. This discrepancy at low $z$ has a straightforward explanation: while we neglected their contribution in our mean-field approximation, we saw with Cayley trees that the likelihood of second and higher neighbors failing is a decreasing function of $z$. 

Fig. \ref{res2}, in turn, plots the empirical distribution of failures $P(F)$ and our analytical estimate thereof (given in Materials and Methods) for ER and BA random networks with $z=8$, for the same values of the financial parameters. Here too, we find that the agreement between the numerical results and the prediction of our mean-field approximation is very good; Fig. \ref{res2} confirms in particular that $P(F)$ is fat-tailed when $p(k)$ is.

The close agreement for these values of the financial parameters (and any other values such that $k^{*}\lesssim 15$, see SI) is remarkable if one contrasts the complexity of the original problem \eqref{repayment} with the extreme simplicity of our mean-field approximation. To us, this conclusion is the main import of our study: once the expression for the critical degree $k^{*}$ as a function of the financial parameters has been obtained, Eq. \eqref{criticaldegree}, analytical results on the systemic risk are not only possible, but also intuitive and straightforward. 


\section{Discussion}

We have considered the effect of financial variables such as interest rates, leverage ratio and financial exposure on the robustness of interbank systems vis-\`a-vis individual shocks. Focusing first on regular networks, we obtained an explicit formula for the critical degree, below which failures begin to propagate through the network. From this, we then showed how to derive a simple but reliable approximation of the expected number of failures and failures distribution in random (and possibly strongly heterogeneous) networks. Interesting extensions of our work could include a non-linear relation between interbank exposure and network degree,\footnote{See  \citep{Soramaki:2007jc} for empirical measures of the correlation between loan size and network degree.} overlapping portfolios, multiple or probabilistic shocks and multiple-period dynamics. 

The highly stylized character of our model notwithstanding, our results shed new light on important aspects of systemic risks, such as the association between contagion and interest rate policy \citep{Freixas:2010uf}. Using plausible values for interest rates, liquidity requirement and leverage, we found critical degrees $k^{*}$ of the order of $5$ to $10$. An empirical analysis of the FedFunds market \citep{Soramaki:2007jc} found a mean degree $z\simeq15$, but almost half of the banks had out-degrees less than $4$, thus vulnerable to contagion. In the 2008 financial crisis, mean degrees in the interbank network declined, increasing systemic risk \citep{Minoiu:2013fa}. While regulators do not directly control the topology of financial networks, it is useful to understand how tools already in place---interest rates, leverage and liquidity requirements---can affect the critical degree.


We observed that our formula \eqref{criticaldegree} for the critical degree $k^*$ reduces to $1/(R-1)$ in the high leverage, low liquidity limit. This limiting value can be expressed as a ``cost-benefit'' rule of thumb, as follows. If bank $i$ lends $l$ to bank $j$ and $j$ does not repay $i$, $i$ will have lost $l$; if on the other hand $j$ makes a successful investment with the money borrowed from $i$ and repays it in full, $j$ will have made a profit profit $(R-1)l$. The critical degree $k^*$ is then just the ratio of the potential loss $l$ to the potential profit $(R-1)l$ of each transaction. Given the great difficulty of the problem of assessing the robustness of actual financial networks, this simple rule of thumb could prove a handy ``order-zero'' approximation. 

What is more, this interpretation establishes a direct link with a seemingly unrelated problem: the condition for the evolution of cooperation, famously investigated by Hamilton \citep{Hamilton:1964wk}. Ref. \citep{Ohtsuki:2006tt} recently extended his insights to graphs and social networks, showing that ``natural selection favours cooperation, if the benefit of the altruistic act, $b$, divided by the cost, $c$, exceeds the average number of neighbours, $k$, which means $b/c > k$''.\footnote{More precisely, the criterion for natural selection to favour cooperation is $b/c > k_{nn}$, where $k_{nn}$ is the mean nearest-neighbor degree, see \citep{Konno:2011wx}.} This simple rule is precisely the same as the one we found for shock propagation in high-leverage, low liquidity interbank networks: the critical degree is given by the ratio of the activities promoting systemic propagation (benefit of cooperation and interbank lending respectively) to the activities inhibiting systemic propagation (cost of cooperation and external profits respectively). This unexpected connection supports the convergence of ecology and finance advocated by Haldane and May after the 2008 crisis \citep{Haldane:2011cg}, and points to a unified perspective on resource sharing in networks.

\section*{Materials and Methods}

\subsection*{Networks}

In this paper we considered three classes of networks: Cayley trees, ER networks and BA networks. They are defined as follows. 

\begin{itemize}
\item
Cayley trees are graphs without loops in which each node is connected to a fixed number of neighbors $k$. Given an (arbitrarily chosen) ``root'' node $i_{0}$, the number of nodes at distance $d$ from $i_{0}$ is $k(k-1)^{d-1}$. 
\item
ER networks are the simplest random networks: given $N$ nodes, each possible edge is included in the network with probability $\phi$, independently from every other edge. When $N\gg1$, this results in a random network with Poisson degree distribution
\begin{eqnarray}
p(k)=e^{-z}\frac{z^{k}}{k!},
\end{eqnarray}
where $z=\phi (N-1)$ is the mean degree. The absence of correlations in such networks entails that the conditional degree distribution---the probability that a node connected to a node with degree $k$ has degree $l$---is just $p(l\vert k)=lp(l)/z$.
\item
BA networks are obtained by means of a stochastic growth process. Starting from a complete graph over (say) $m$ initial nodes, each new node is added to $m$ existing nodes with a probability that is proportional to the number of links that the existing nodes already have. In the large time limit, this process defines a correlated random network with (conditional) degree distribution \citep{Fotouhi:2013cr}
\begin{eqnarray}
p(k)&=&\frac{2m(m+1)}{k(k+1)(k+2)},\\
p(l\vert k)&=&\frac{m}{kl}\left(\frac{k+2}{l+1}-{2m+2\choose m+1}\frac{{k+l-z\choose l-m}}{{k+l+2\choose l}}\right),
\end{eqnarray}
where $k\geq m$; the mean degree is given by $z=2m$.

\end{itemize}

\subsection*{Failures distribution}
Consider a random network with degree distribution $p(k)$ and conditional degree distribution $p(k\vert l)$. Suppose that the shocked bank $i_{0}$ has degree $k$, and let $q(k)=\sum_{l=1}^{k^{*}}p(l\vert k)$ be the probability that a neighbor of the shocked bank $i_{0}$ has a subcritical degree. According to our ``mean-field'' assumption, the probability that $F$ first neighbors of $i_{0}$ fail is given by the probability $q(k)^{F}$ that $F$ neighbors have subcritical degree, times the probability $[1-q(k)]^{k-F}$ that $k-F$ neighbors have supercritical degree, times the number of choices of $F$ failing neighbors among $k$. Weighing this by the probability $p(k)$ that $i_{0}$ has $k$ neighbors, we arrive at 
\begin{eqnarray}\label{proba2}
P(F)=\sum_{k\geq F}p(k){k \choose F}q(k)^{F}[1-q(k)]^{k-F}+(\textrm{contribution of higher neighbors}).
\end{eqnarray}
The expected number of failures \eqref{failures} is then obtained by evaluating $\langle F\rangle=\sum_{F\geq1}FP(F)$ (see SI). Note that expression \eqref{proba2} is strongly reminiscent of the classical theory of percolation on complex networks, where one shows \citep{Cohen:2001hf} that the degree distributions $p'(k')$ after the removal of a fraction $q$ of the nodes is given in terms of the old degree distribution $p(k)$ by $p'(k')=\sum_{k\geq k'}p(k){k \choose k'}q^{k}(1-q)^{k'-k}$. This is no surprise: the whole point of our mean-field approximation is to reduce a dynamical problem (computation of repayments) to a topological one (failure depends on degree only).


\subsection*{Large $F$ asymptotics}

Let us now consider the limit of \eqref{proba2} when $F\gg1$ (hence for shocked banks with degree $k\gg1$), assuming that $q(k)$ becomes independent of $k$ in this limit. (This amounts to saying that correlations between the degrees $k$ and $l$ of adjacent nodes become immaterial when $\vert k-l\vert\gg1$; this holds for both ER and BA networks.) Let us consider Poisson-distributed and power-law distributed networks separately (see SI for details).

\begin{itemize}
\item
\emph{Poisson networks}. Given the Poisson degree distribution $p(k)=e^{-z}z^{k}/k!$, resumming \eqref{proba2} is straighforward and gives 
\be
P(F)= e^{-zq}\frac{(zq)^{F}}{F!}.
\ee
Thus, in Poisson distribued networks, the failures distribution is Poissonian with mean $zq$. 
\item
\emph{Scalefree networks}. For scalefree networks we observe that, when $\gamma$ is an integer and for sufficiently large $k$, the Pochhammer symbol $(k)_{\gamma}=k(k+1)\dots (k+\gamma-1)$ can be substituted to $k^{\gamma}$ in the degree distribution $p(k)\sim 1/k^{\gamma}$. This allows to perform the sum \eqref{proba2} explicitly, yielding
\be\label{failuresSF}
P(F)\sim \frac{q^{F}\,{}_2 F_1(F,F+1;F+\gamma;1-q)}{(F)_{\gamma}} \sim \frac{q^{\gamma-1}}{F^{\gamma}},
\ee
where in the second step we used the asymptotics of the Gauss hypergeometric function ${}_2 F_1$ for large parameters \citep{Temme:2003es}. Analytical continuation in $\gamma$ then shows that $P(F)$ is scalefree with exponent $\gamma$ also for non-integer $\gamma$. 
\end{itemize}
In both cases, the tail of $P(F)$ has the same nature (Poisson or power-law) as the degree distribution itself.

\section*{Acknowledgements}
We thank the participants and staff of the 2013 Santa Fe Institute Complex Systems Summer School (where this project was initiated) for a very stimulating experience. We are especially indebted to T. J. Carter, R. Martinez and M. M. King and for their help in the early stages of this research, and to B. Vaitla for drawing our attention to Ref. \citep{Ohtsuki:2006tt}. Useful discussions with V. Bonzom are gratefully acknowledged. Research at the Perimeter Institute is supported in part by the Government of Canada through Industry Canada and by the Province of Ontario through the Ministry of Research and Innovation.

\bibliographystyle{abbrvnat}
\bibliography{library}

\newpage 

\appendix

\setcounter{figure}{0} \renewcommand{\thefigure}{S\arabic{figure}}

\begin{center}
\Large{Mapping systemic risk: critical degree and failures distribution in financial networks\\ \bigskip Supplementary Information}
\end{center}

\section{Computation of the critical degrees $k_{(1)}^{*}$ and $k_{(2)}^{*}$}

On a regular graph with degree $k$, the liquid assets $\lambda_{i}$, senior liabilities $s_{i}$ and investment profit $\rho_{i}$ of a bank $i$ given assumptions $(i-iv)$ are all proportional to $k$, given respectively by
\begin{eqnarray}
\lambda_{i}&=&\frac{f}{1-f}\, k,\\
s_{i}&=&\frac{f-\Lambda}{1-f}\, k,\\
\rho_{i}&=&\frac{(R_{i}-1)(1-\Lambda)}{1-f}\, k,
\end{eqnarray}
with $R_{i}=R$ if $i\neq i_{0}$ and $R_{i}=0$ if $i=i_{0}$. The repayment equation \eqref{repayment} can therefore be written as 
\be\label{regrepay}
x_{i}=\left[\min\Big\{\frac{(R_{i}-1)(1-\Lambda)+\Lambda}{1-f}\,k+\sum_{j\leftrightarrow i}\frac{x_{j}}{k},rk \Big\}\right]^{+}
\ee
where the sum ranges over the neighbors $j$ of $i$. Moreover, since in a Cayley tree all banks $i$ at the same distance $d$ from the shocked bank $i_{0}$ are equivalent, their repayments $x_{i}$ must take a common value $x_{(d)}$, with $0\leq x_{(1)}< x_{(2)} <\cdots\leq rk$. Thus \eqref{regrepay} becomes
\be\label{regrepay2}
x_{(0)}=\left[\frac{-1+2\Lambda}{1-f}\,k+x_{(1)}\right]^{+}
\ee
for the shocked bank itself and
\be\label{regrepay3}
x_{(d)}=\frac{(R-1)(1-\Lambda)+\Lambda}{1-f}\,k+\frac{x_{(d-1)}}{k}+(k-1)\frac{x_{(d+1)}}{k},\quad\quad d\geq1
\ee
for its first, second and higher neighbors. From \eqref{regrepay2} and \eqref{regrepay3} it is easy to compute the first few critical degrees $k_{(d)}^{*}$. 

By definition the first critical degree $k_{(1)}^{*}$ is such that 
\begin{itemize}
\item
only the shocked bank defaults, viz.
\be
x_{(d)}=rk_{(1)}^{*}\quad\textrm{for}\quad d\geq1,
\ee
\item
the first neighbors of $i_{0}$ are critical, viz.
\be\label{regrepay4}
x_{(1)}=\frac{(R-1)(1-\Lambda)+\Lambda}{1-f}\,k_{(1)}^{*}+\frac{x_{(0)}}{k_{(1)}^{*}}+(k_{(1)}^{*}-1)r=rk_{(1)}^{*}
\ee 
\end{itemize}
Solving \eqref{regrepay2} and \eqref{regrepay4} for $k_{(1)}^{*}$ yields
\be
k_{(1)}^{*}=\frac{r(1-f)-[r(1-f)+2\Lambda-1]^+}{(R-1)(1-\Lambda)+\Lambda}.
\ee
Similarly, the second critical degree $k_{(2)}^{*}$ corresponds to the situation where
\begin{itemize}
\item
only the shocked bank and its first neighbors default, viz.
\be
x_{(d)}=rk_{(2)}^{*}\quad\textrm{for}\quad d\geq2,
\ee
\item
the second neighbors of $i_{0}$ are critical, viz.
\be
x_{(2)}=\frac{(R-1)(1-\Lambda)+\Lambda}{1-f}\,k_{(2)}^{*}+\frac{x_{(1)}}{k_{(2)}^{*}}+(k_{(2)}^{*}-1)r=rk_{(2)}^{*}
\ee  
\end{itemize}
This gives (assuming $r<(1-2\Lambda)/(1-f)$, so that $x_{(0)}=0$):
\be\label{critical2}
k_{(2)}^{*}=\frac{1}{2}\left(\sqrt{1+\frac{4r(1-f)}{{(R-1)(1-\Lambda)+\Lambda}}}-1\right).
\ee
Observe that $k_{(2)}^{*}=\mathcal{O}(k_{(1)}^{* 1/2})$, hence the second critical degree $k_{(2)}^{*}$ grows much more slowly than the first critical degree $k_{(1)}^{*}$. This suggests that, except in the extreme case where $k_{(1)}^{*}\gg1$ (which can happen only if  $R\rightarrow1$ and $f,\Lambda\rightarrow0$), the direct propagation of failures to second and higher neighbors of the shocked bank is excluded in our model (within assumptions $(i-v)$).


%
%
\section{Proof of Eq. \eqref{failures}}


Given the estimate \eqref{proba2}, the expected number of failures $\langle F\rangle=\sum_{F\geq1}FP(F)$ can be written as
\begin{eqnarray}\label{appendix1}
\langle F\rangle=\sum_{k\geq1}p(k)\sum_{F=0}^{k}F{k \choose F}q(k)^{F}[1-q(k)]^{k-F}.
\end{eqnarray}
Now, using Newton's binomial formula it is easy to show that, for any two numbers $X$ and $Y$, 
\be\label{newton}
\sum_{F=0}^{k}F{k \choose F}X^{F}Y^{k-F}=kX(X+Y)^{k-1}.
\ee
Using relation \eqref{newton} with $X=q(k)$ and $Y=1-q(k)$ in Eq. \eqref{appendix1} gives 
\be
\langle F\rangle=\sum_{k\geq1}kp(k)q(k).
\ee
Note that, for uncorrelated networks (for which $p(l\vert k)=lp(l)/z$), $q(k)$ is independent of $k$, hence $\langle F\rangle=zq$.

\section{Proof of Eq. \eqref{failuresSF}}

The Gauss hypergeometric function with parameters $(a,b,c)$ is defined as the series
\be
{}_2 F_1(a,b;c; \zeta)=\sum_{n=0}^{\infty}\frac{(a)_{n}(b)_{n}}{(c)_{n}}\frac{\zeta ^{n}}{n!}
\ee
where $(m)_{n}=m(m+1)\dots(m+n-1)$ is the Pochhammer symbol and $\zeta$ is a complex number. (Note the potentially confusing notation: ${}_2 F_1$ is the Gauss hypergeometric function, and $F$ is the number of failures induced by a shock.) Its asymptotic behavior in the limit of large $c$ parameter is given by Watson's expansion \citep[p. 397]{Olver:2010vy}, yielding in particular
\be\label{watson}
\lim_{F\rightarrow\infty}{}_2 F_1(a,b;c+F;\zeta)=1.
\ee
Using the connection formula
\be
{}_2 F_1(a,b;c;\zeta)=(1-\zeta)^{c-a-b}{}_2 F_1(c-a,c-b;c;\zeta),
\ee
this gives
\be\label{asym}
\lim_{F\rightarrow\infty}{}_2 F_1(a+F,b+F;c+F;\zeta)=(1-\zeta)^{c-a-b}.
\ee

Consider the expression \eqref{proba2} for the failures distribution, assuming that the number of failures $F$ is large enough (so that $q(k)$ has a constant value $q$), and plug in a power-law degree distribution of the form $p(k)\sim1/(k)_{\gamma}$: 
\begin{eqnarray}
P(F)&\sim& q^{F}\sum_{k\geq F}\frac{1}{(k)_{\gamma}}{k \choose F}(1-q)^{k-F}=q^{F}\sum_{n\geq 0}\frac{1}{(F+n)_{\gamma}}{F+n \choose F}(1-q)^{n}.
\end{eqnarray}
Rewriting
\be
{F+n \choose F}=\frac{(F+1)_{n}}{n!}
\ee
and
\be
\frac{1}{(F+n)_{\gamma}}=\frac{1}{(F)_{\gamma}}\frac{(F)_{n}}{(F+\gamma)_{n}},
\ee
we get
\be
P(F)\sim \frac{q^{F}\,{}_2 F_1(F,1+F;\gamma+F;1-q)}{(F)_{\gamma}}.
\ee
Using the asymptotic formula \eqref{asym}, we arrive at
\be
P(F)\sim\frac{q^{\gamma-1}}{F^{\gamma}}.
\ee

\newpage
\section{Mean number of failures: varying interest rate and leverage ratio}

In sec. \ref{num} we studied the mean number of failures induced by a single shock within two network ensembles: Erd\"os-R\'enyi random networks and Barab\'asi-Albert scalefree networks. Specifically, we compared our analytical estimate \eqref{failures} with numerical results obtained by averaging over $10^{4}$ networks for each value of the mean degree $z$.   

Fig. S1 presents further results showing the effect of varying the leverage ratio $\Lambda$ and the external interest rate $R$. While the agreement between theory (circles) and numerics (dots) remains qualitatively good for all considered values, we observe that systematic discrepancies---notably at low $z$---arise when $\Lambda$ and $R$ get large. This corresponds to regimes where the critical degree $k^{*} \gtrsim 15$. In such regimes, second and higher neighbors of the shocked bank are likely to fail, as the second critical degree $k^{*}_{(2)}$ becomes significantly larger than zero. It is an interesting challenge to extend our mean-field approximation so as to capture such higher-neighbor effects.

\bigskip

\begin{figure}[h!]
\centering
\includegraphics[scale=.9]{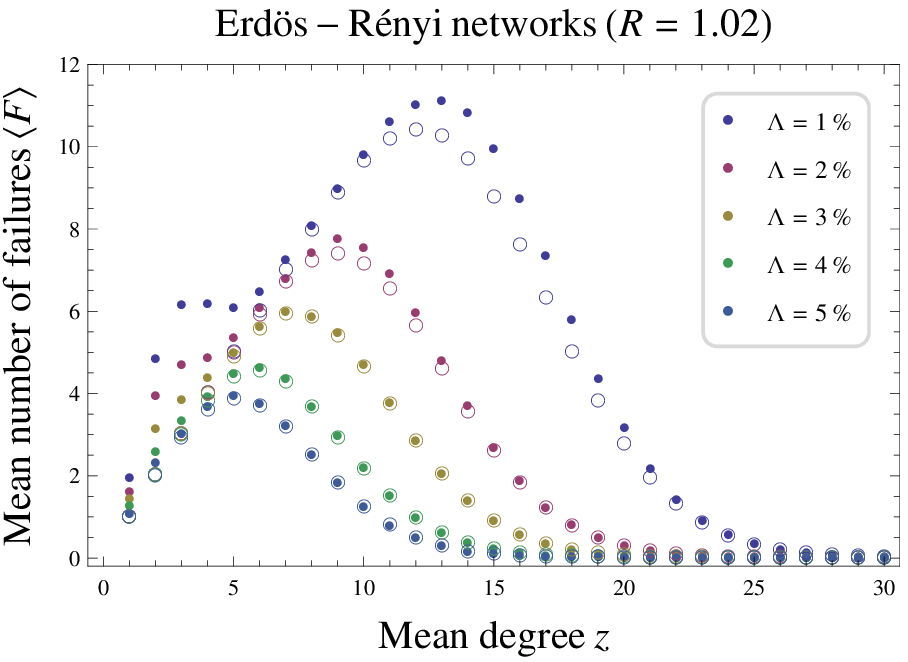} \hfill
    \includegraphics[scale=.9]{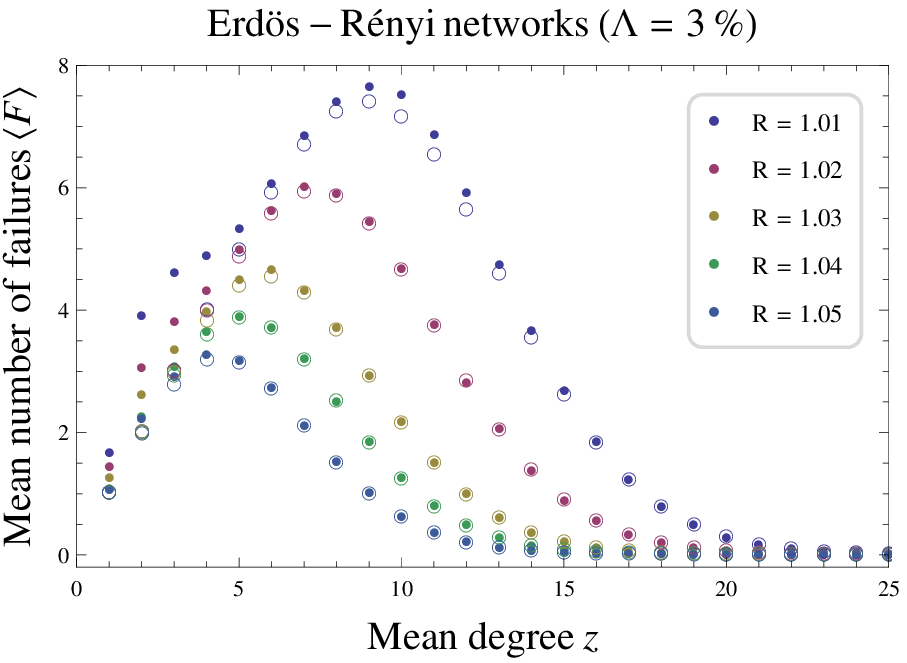}    
        
    \bigskip
    \bigskip
    
         \includegraphics[scale=.9]{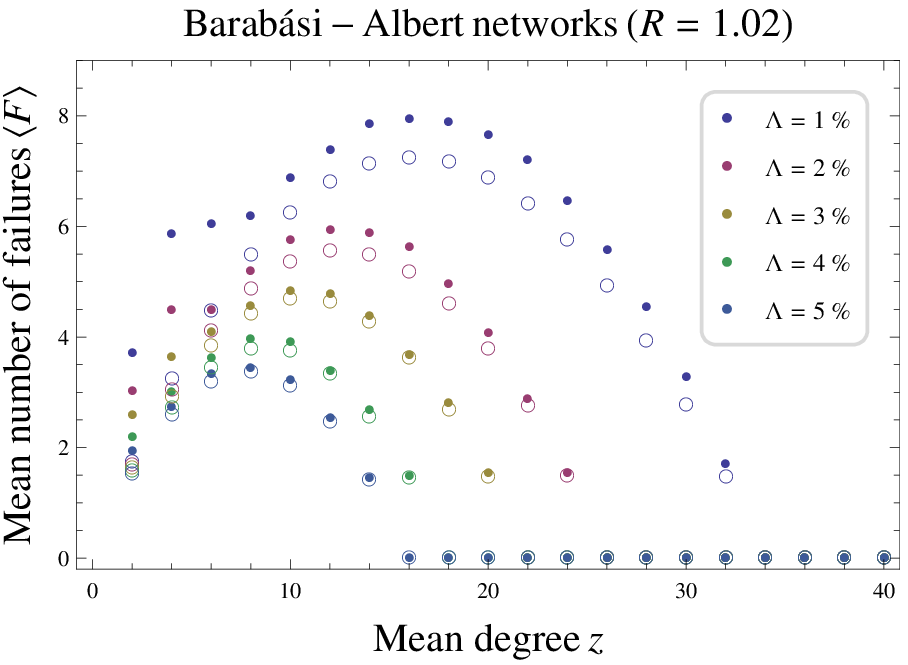}\hfill
    \includegraphics[scale=.9]{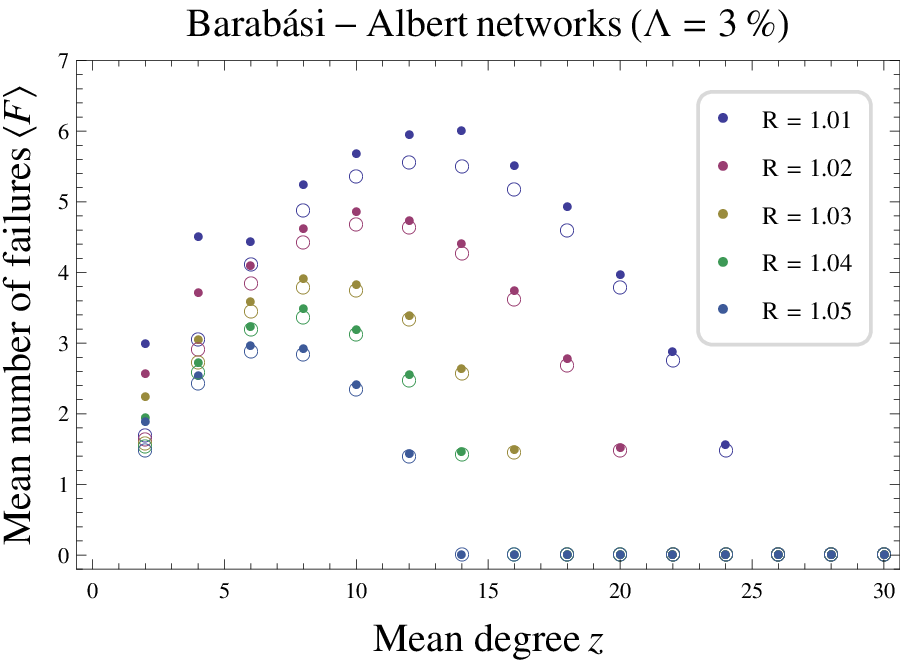}

               \caption{Mean number of failures in ER (top) and BA (bottom) networks as a function of mean degree $z$, for $r=1.01$ and $f=50\%$. The circles represent our mean-field estimate \eqref{failures}, the dots represent the results of numerical averages over $10^{4}$ networks with $N=100$ banks. In the left column we vary $\Lambda$ at fixed $R=1.02$; in the right column we vary $R$ at fixed $\Lambda=3\%$ (right).}
\end{figure}

\end{document}